\documentclass[journal,twoside]{IEEEtran}
\usepackage[T1]{fontenc}
\usepackage{cite}
\usepackage{amsmath,amssymb,amsfonts}
\usepackage{algorithmic}
\pdfoutput=1
\usepackage[]{graphicx}
\usepackage[mathlines,switch]{lineno}
\usepackage{multicol,multirow}
\usepackage{epsfig}
\usepackage{subfigure}
\usepackage{lineno}
\usepackage{TUSON}

\title{CrossRAFT: Cross-Domain Complex-Valued Feature Extraction for Ultrasound Motion Estimation}
\author{Yang Leng, \IEEEmembership{Student Member, IEEE}, Yuchen Tang, \IEEEmembership{Graduate Student Member, IEEE}, Kai-Hang Yiu, Yik-Chung Wu, \IEEEmembership{Senior Member, IEEE}, and Wei-Ning Lee \IEEEmembership{Senior Member, IEEE}
\thanks{This work was supported by the Hong Kong General
Research Fund under Grant 17205022.}
\thanks{Yang Leng, Yuchen Tang, and Yik-Chung Wu are with the Department of Electrical and Computer Engineering, The University of Hong Kong, Hong Kong. (e-mail: u3008920@connect.hku.hk, tyc2017@connect.hku.hk, ycwu@eee.hku.hk). Kai-Hang Yiu is with The University of Hong Kong-Shenzhen Hospital and Department of Medicine, The University of Hong Kong, Hong Kong (e-mail: khkyiu@hku.hk). Wei-Ning Lee is with the Department of Electrical and Computer Engineering, and also with the School of Biomedical Engineering, The University of Hong Kong, Hong Kong. (e-mail: wnlee@hku.hk).}}

\IEEEaftertitletext{\GA{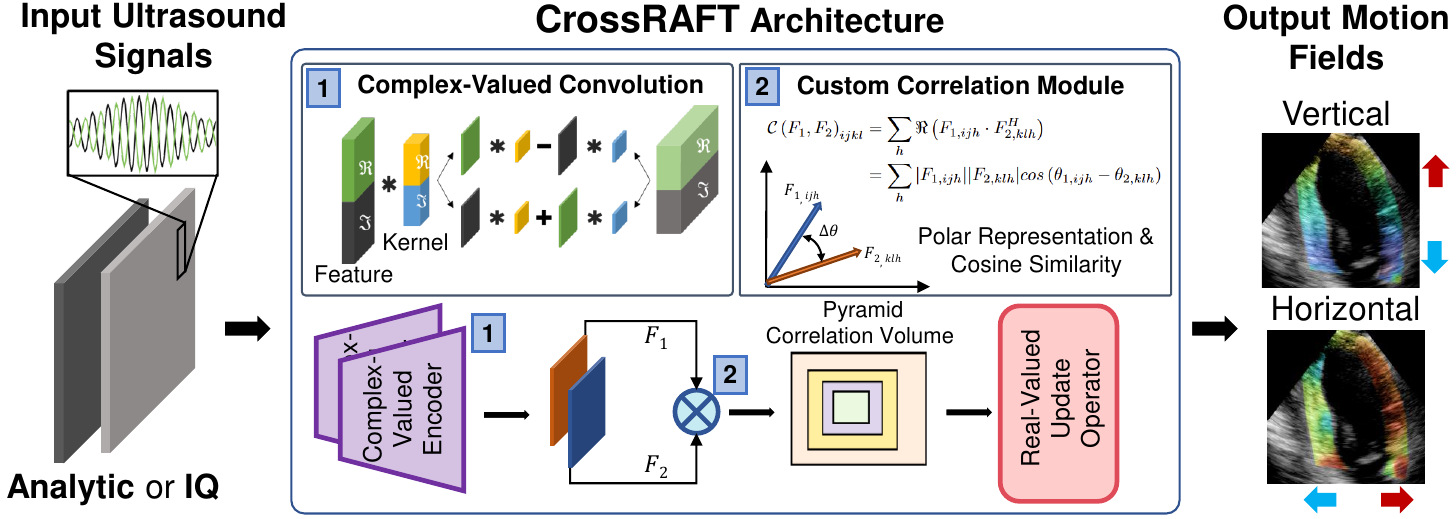}
\Abstract{\begin{abstract}
Accurate multi-dimensional motion estimation is fundamental to broad biomedical ultrasound applications, primarily performed using real-valued radio-frequency (RF), analytic, and in-phase and quadrature (IQ) signals. Real-valued RF and analytic signals offer high subsample accuracy but are computationally intensive. Regardless of the signal domain used, traditional algorithms suffer from interframe decorrelation, resolution tradeoffs, and a lack of a carrier and low sampling in the lateral direction. Emerging deep learning (DL) models demonstrate improved performance and faster inference. However, existing architectures are real-valued and disrupt inherent phase coupling crucial for accurate motion estimation by treating complex components as independent channels. 
To address this, we propose CrossRAFT, an end-to-end trainable network designed for subsample motion estimation directly from complex-valued ultrasound signals. Built on the Recurrent All-Pairs Field Transform (RAFT) architecture, CrossRAFT features complex-valued encoders and a custom correlation module to explicitly extract and use phase information. The model is pre-trained via supervised learning and subsequently adapted to diverse tissue dynamics via bidirectional unsupervised learning. 
We evaluated CrossRAFT across real RF, analytic, and IQ signal presentations on simulated phantoms, synthetic echocardiographic, and private \textit {in vivo} echocardiographic datasets. Experimental results show that CrossRAFT estimates displacements accurately under complex tissue motion and in the challenging lateral direction. Compared to real RF-based RAFT, analytic-based CrossRAFT reduces vertical and horizontal displacement errors by 69\% and 51\% on the synthetic echocardiographic data, respectively. 
These findings demonstrate that leveraging phase information via complex‑valued neural networks provides a powerful framework for ultrasound motion estimation, showing potential for clinical elastography and functional imaging.
\hblne
\end{abstract}}
\vspace{1\baselineskip}\vspace*{-1pt}
}

\begin{document}

\maketitle

\Highlights[1]{We propose CrossRAFT, a cross‑domain framework that explicitly leverages complex‑valued ultrasound signals (analytic and IQ) for motion estimation.}

\Highlights[2]{CrossRAFT with analytic input effectively reduces vertical and horizontal displacement errors by 69\% and 51\%, respectively, on synthetic cardiac data versus RF‑based RAFT.}

\Highlights[3]{This work establishes a new paradigm that leverages complex‑valued neural networks for accurate motion estimation in elastography and functional imaging.}


\PrintHighlights

\section{Introduction}
\IEEEPARstart{M}{otion} estimation in ultrasound imaging aims to quantify tissue and blood motion from sequential acquisitions at sufficient frame rates. It provides essential kinematic parameters for biomechanical analysis, such as Young's or shear modulus in elasticity imaging~\cite{ophir1991elastography}, myocardial deformation in tissue Doppler imaging and strain (rate) imaging~\cite{mcdicken1992colour,heimdal1998real}, and shear stress or vorticity in blood flow imaging~\cite{bohs1991novel}. Motion estimation is essentially finding similarity between sequentially acquired backscattered signals, whose constructive and destructive inferences produce speckle patterns shown in a B-mode image. These patterns serve as intrinsic natural tissue markers~\cite{insana1990speckle} that encode motion-induced changes in backscattered signals. Motion estimation in ultrasound imaging is known as ultrasound speckle tracking.

A variety of speckle tracking algorithms have been developed. Among them, window-based methods, particularly with normalized cross-correlation (NCC) applied to radio-frequency (RF) or envelope-detected data, are robust and form the historical foundation. These methods operate by defining a reference window in the pre-deformed frame and searching for the region of highest similarity within a search range in the post-deformed frame to estimate local displacement vectors. RF-based NCC methods achieve accurate axial displacement estimation for small deformation due to the carrier frequency and phase information retained in the RF signal~\cite{hein1993current}, whereas envelope-based methods are often suitable for tracking tissues with large deformation~\cite{ma2013comparison}. 

NCC algorithms are also extended into the frequency domain~\cite{o1991measurement}, giving rise to a family of phase‑zero crossing techniques~\cite{pesavento1999baseband, o1993quantitative, o2002internal, lubinski1999adaptive, lubinski2002speckle,kaluzynski2001strain,huang2007p4b,huang2009phase,huang2010analytic}. These methods estimate displacements by detecting the zero crossing of the phase difference between complex-valued signals from pre‑ and post‑deformation frames, achieving high axial precision. However, NCC-based approaches are sensitive to noise and signal decorrelation, and their performance is notably limited in the lateral dimension due to the lack of phase information and lower spatial resolution~\cite{hein1993current}. 

To further tackle the challenge in lateral motion estimation, RF‑based lateral motion estimation techniques have been developed using interpolation~\cite{konofagou1998new, techavipoo2004estimation, luo2009effects,kothawala2017spatial} and augmentation~\cite{selladurai2018strategies} strategies. Meanwhile, several analytic-based lateral motion estimators have also been explored, including phase‑sensitive lateral motion estimators~\cite{hasegawa2009analytic} and synthetic lateral phase (SLP) techniques~\cite{chen2004lateralSLP,huang2010analytic,el2011brainSLP}. These SLP‑based methods operate directly on the analytic signal by preserving both its positive and negative frequency components in the lateral spectrum, thereby enhancing the accuracy of lateral motion estimation. Notably, Huang \textit{et al.}~\cite{huang2010analytic} incorporated this SLP estimator with a frequency‑domain filter, leading to a 46\% reduction in the variance of lateral motion estimates.

In parallel, several optimization-based methods~\cite{khamis2016optimazation,GLUE2017global,Muk2020log,SOUL2021combining} have been proposed to formulate ultrasound motion estimation as the minimization of an energy function. For example, one study~\cite{SOUL2021combining} combined data fidelity, mechanical consistency, and smoothness constraints within a unified cost function for two-dimensional (2D) axial and lateral speckle tracking in ultrasound elastography, achieving a high signal-to-noise ratio (SNR) of 88.17 in a breast elastography phantom. However, this method is highly user-dependent and requires careful manual tuning of regularization weights to generate strain images, thus limiting their practical efficiency and clinical adoption. Furthermore, it has predominantly been applied in quasi‑static elastography and has not yet been validated for tracking tissues with spontaneous dynamics, such as the myocardium.

Deep learning (DL) -based methods provide a new paradigm for ultrasound motion estimation through their capability of learning complex motion representations directly from data as well as high computational efficiency. Models originally developed for optical flow in computer vision fields, such as Flow-Net~\cite{FlyingChairs2015flownet}, PWC-Net~\cite{sun2018pwc}, and IRR-Net~\cite{Hur_2019_CVPR}, have demonstrated superior capability in capturing detailed motion dynamics from natural images. Their adaptations to ultrasound~\cite{flownet2018ultrasound,tehrani2020mpwc,tehrani2021mpwc++, tehrani2022bimpwc} have also achieved performance comparable to that of traditional speckle tracking methods~\cite{mirzaei2019overwind}. Generally, these adaptions involve a two-stage training pipeline: supervised pre-training on large computer vision datasets (e.g., FlyingChairs~\cite{FlyingChairs2015flownet}) to acquire the ability to capture motion, followed by fine-tuning on the concatenation of various ultrasound signals, including RF, envelope, and B-mode.  
Such a multi‑input design helps the model extract useful features from diverse ultrasound formats. However, this pipeline suffers from a fundamental domain gap: natural images and ultrasound data differ significantly in physical properties, appearance, and motion statistics. The former are often rigid with large displacements, whereas tissue motion is often smooth and deformable. Hence, features learned on natural images may transfer suboptimally, and any architectural modification for ultrasound signals requires retraining on large‑scale computer vision datasets.

Recent progress in DL-based optical flow has introduced Recurrent All-Pairs Field Transform (RAFT)~\cite{teed2020raft}, an advanced architecture distinguished by its robust iterative refinement mechanism based on a dense correlation volume. This design makes RAFT a particularly promising candidate for adaptation to motion estimation in ultrasound imaging. Initial success, such as RAFT-USENet~\cite{majumder2025useraft}, has applied RAFT to estimate tissue deformation under quasi-static compression. RAFT-USENet~\cite{majumder2025useraft} follows a multi‑input strategy that concatenates beamformed RF signals, their Hilbert-transformed imaginary parts, and envelope data as model input of a real‑valued network. RAFT-USENet also incorporates tissue incompressibility into the loss function, but such a constraint applied to a 2D imaging plane is an incomplete description of three‑dimensional deformation. Evaluated on a simulated phantom dataset~\cite{tehrani2020mpwc}, this approach achieved mean absolute errors of $2.82 \mu m$ and $8.61 \mu m$  across all strain levels in the axial and lateral directions, respectively.

However, the aforementioned DL models operate on real-valued neural networks. They inherently disrupt the natural correlation between the real and imaginary parts of an analytic signal by treating RF signal, imaginary part, envelope, and B‑mode images as independent real-valued channels. In fact, these components can all be presented in the analytic signal, which is derived from the RF signal via the Hilbert transform and separates local amplitude from instantaneous phase at every sample point~\cite{wachinger20122d}. Unlike in-phase and quadrature (IQ) data, which is a complex baseband representation, the analytic signal offers a unique advantage because it preserves the original RF carrier frequency and sampling resolution, enabling high‑precision motion estimation that is insusceptible to amplitude variations. 

Nevertheless, existing DL models cannot directly process complex‑valued inputs-whether IQ or analytic —as they are built upon real‑valued operations. This gap motivates the exploration of complex‑valued neural networks (CVNNs) for motion estimation from complex-valued ultrasound data. Yet the target displacement fields are real‑valued, which necessitates a cross‑domain architecture that bridges the complex-valued input space and the real-valued output space. 

To this end, we propose CrossRAFT, a cross-domain RAFT framework that explicitly leverages complex‑valued ultrasound signal representations for motion estimation. This is an end-to-end trainable network dedicated for motion estimation directly from complex-valued signal representations. The framework pioneers the use of CVNNs to extract critical phase information from input data and incorporates a correlation module specifically designed for bridging complex- and real-valued domains. By enabling end‑to‑end learning from complex‑valued inputs to real‑valued displacement fields, this work fills the gap in DL‑based motion estimation from complex ultrasound signals. We perform a systematic comparison of motion estimation performance across three different types of ultrasound signals: beamformed RF signals, IQ-demodulation data, and analytic signals, on publicly available simulated phantom~\cite{tehrani2020mpwc} and echocardiographic data~\cite{burman2024COLE}, as well as private \textit{in vivo} echocardiographic data. 

The remainder of this paper is organized as follows: Section \ref{sec_method} introduce details of the proposed cross-domain framework and its architectural components. Section \ref{sec_3} describes the experimental setup, datasets, and evaluation metrics. Sections~\ref{sec_result} and \ref{sec_dis} present and discuss the results, and Section~\ref{sec_conclu} concludes the paper with final remarks and future directions.

\section{Methods}
\label{sec_method}
\subsection{Problem Formulation}
Ultrasound motion estimation involves estimating a dense displacement vector field that maps points in a source frame to a target frame. Consider a pair of ultrasound frames, denoted as $\mathbf{I_1}$ and $\mathbf{I_2}$. We define $H$ and $W$ as the height and width of each frame, respectively. The objective is to find a function $\mathcal{F}$ that estimates the displacement field $\mathbf{f}_{1 \rightarrow2} = \mathcal{F}\left(\mathbf{I_1},\mathbf{I_2};\Theta \right)$,
where $\Theta$ represents the parameters of $\mathcal{F}$, and $\mathbf{f}_{1 \rightarrow2}\left(u, v\right)$ is the 2D displacement vector $(d_x, d_y)$ that maps the spatial point at location $(u, v)$ in $\mathbf{I_1}$ to its corresponding position in $\mathbf{I_2}$.
Under the assumption of brightness constancy in optical flow, %which is adapted for ultrasound as "speckle pattern constancy" or "phase consistency" for small displacements, 
the underlying signal should satisfy: \begin{equation}
    \mathbf{I}_2\left(\mathbf{x+\mathbf{f(x)}}\right) \approx \mathbf{I}_1, 
\end{equation}
where $\mathbf{x}=(u, v)$ denotes a pixel location, and $\mathbf{f(x)} = (d_x, d_y)$ is the displacement at that location. However, this idealized model is complicated by several intrinsic factors in ultrasound imaging, such as speckle decorrelation and noise. Therefore, the problem is reformulated as finding the most probable displacement field $\mathbf{f}_{1\rightarrow2}$ given the ultrasound data pair $(\mathbf{I_1}, \mathbf{I_2})$. This can be cast as a maximum a posteriori (MAP) estimation problem:
\begin{equation}\hat{\mathbf{f}}_{1\to2}=\arg\max_{\mathbf{f}}P(\mathbf{f}\mid\mathbf{I}_1,\mathbf{I}_2).\end{equation}
Using Bayes' theorem, this is proportional to:
\begin{equation}\hat{\mathbf{f}}_{1\to2}=\arg\max_{\mathbf{f}}P(\mathbf{I}_1,\mathbf{I}_2\mid\mathbf{f}) P(\mathbf{f}),\end{equation}
where $P(\mathbf{I}_1,\mathbf{I}_2\mid\mathbf{f})$ is the likelihood function, modeling the probability of pre- and post-deformed pair given a displacement field. $P(\mathbf{f})$ is the prior term, which contains assumptions about the smoothness and regularity of the motion field.

Directly solving this MAP inference is computationally intractable for dense fields. Inspired by RAFT's success in natural images, which implicitly learns both likelihood and prior through its iterative refinement process, we adopt a similar DL paradigm. The central task of this framework is to learn the model parameters $\Theta$ that minimize a supervised loss $\mathcal{L}$ over a dataset of $N$ ground-truth displacement fields ($\mathbf{f}_{gt}$):
\begin{equation}\Theta^*=\arg\min_{\Theta}\frac{1}{N}\sum_{i=1}^{N}\mathcal{L}_\text{sp}\left(\mathcal{F}(\mathbf{I}_{1}^{(i)},\mathbf{I}_{2}^{(i)};\Theta),\mathbf{f}_{\mathrm{gt}}^{(i)}\right).\end{equation}

However, ground-truth displacement fields are unavailable in clinical settings, so an unsupervised learning paradigm becomes essential. Hence, instead of directly regressing $\mathbf{f}_{\mathrm{gt}}$, we adopt a photometric reconstruction loss, which enforces the consistency between the source image and the warped target image. The estimated displacement field $\mathbf{f} = \mathcal{F}(\mathbf{I_1}, \mathbf{I_2}; {\Theta})$ is then used to spatially transform $\mathbf{I_2}$ into $\mathbf{I_2}(\mathbf{x}+\mathbf{f(\mathbf{x})}) \triangleq \mathbf{I_2}\circ\mathbf{f}$, such that $\mathbf{I_2} \circ \mathbf{f} \approx \mathbf{I_1}$. The unsupervised cost function $\mathcal{L}_\text{unsp}$ typically integrates a similarity metric $\mathcal{S}$ to measure the similarity between $\mathbf{I_1}$ and the warped image, and a regularization term $\mathcal{R}$ to impose smoothness on $\mathbf{f}$ as follows:
\begin{small}
\begin{align}
    \mathcal{L}_\text{unsp}(\mathbf{I_1}, \mathbf{I_2}; {\Theta})=\mathcal{S}(\mathbf{I_1},\mathbf{I_2}\circ\mathcal{F}(\mathbf{I_1}, \mathbf{I_2}; {\Theta})) + \lambda\mathcal{R}(\mathcal{F}(\mathbf{I_1}, \mathbf{I_2}; {\Theta})),
\end{align} 
\end{small}where $\lambda$ controls the regularization strength, and a candidate for the similarity measure $\mathcal{S}$ is the structural similarity index measure (SSIM). The transition from supervised to unsupervised learning enables the model to capture the underlying motion characteristics and adapt to diverse tissue motion patterns, without dependence on annotated ground-truth data.

\subsection{Model Architecture}
\begin{figure*}[ht]
    
    \centering
    \includegraphics[scale=0.61]{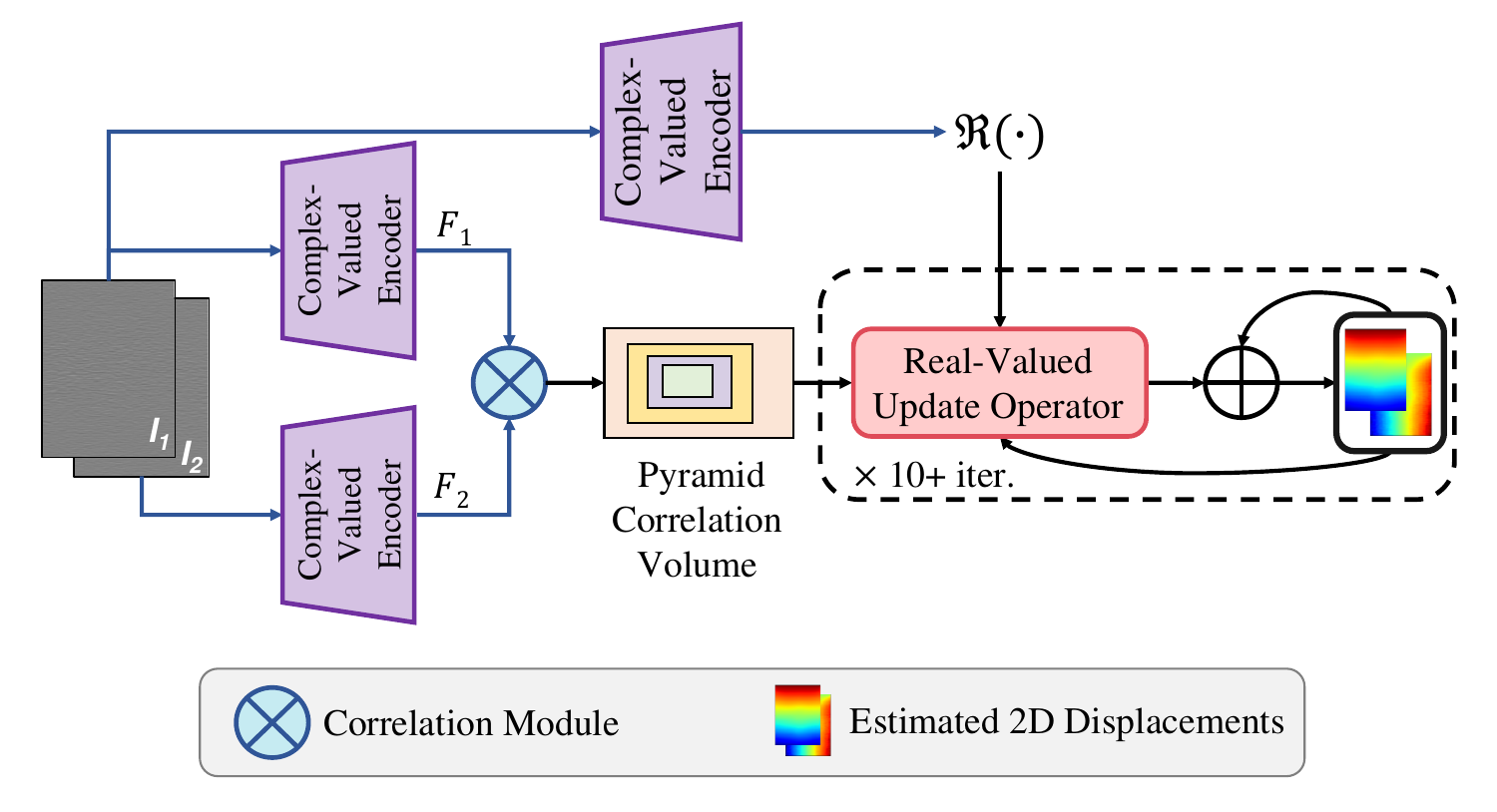}
    \caption{Proposed CrossRAFT architecture (developed based on original RAFT network~\cite{teed2020raft}) to estimate axial and lateral displacements from complex-valued ultrasound data. Blue and black arrows indicate complex‑valued and real‑valued features, respectively.}
    \label{Fig_CrossRAFT}
    \end{figure*}

Fig.~\ref{Fig_CrossRAFT} illustrates the architecture of our proposed CrossRAFT, which comprises three parts: (1) Complex-valued encoders for complex-valued data, extracting phase information through complex-valued convolutions; (2) A pixel-to-pixel correlation module, computing real-valued correlation volumes from complex-valued features to locate matched pixels between frames; (3) A real-valued recurrent update operator that iteratively refines motion fields with local pyramid correlation volumes.
The input to CrossRAFT comprises a pair of pre-deformed and post-deformed analytic or IQ frames, $I_1, I_2 \in \mathbb{C}^{3 \times H_o \times W_o}$, where $H_o = H/4$ and $W_o = W/2$. They are obtained by replicating a single channel 3 times across the channel dimension. Normalization is performed on the amplitude of complex-valued signals for all datasets. A dense motion field $\hat{\mathbf{f}} \in \mathbb{R}^{2 \times H \times W}$ is then estimated, where $2$ denotes displacement components in the axial and lateral directions. The smooth cross-domain transition from the complex-valued ultrasound signal domain to the real-valued displacement domain is realized by the pixel-to-pixel correlation module. Next, we elaborate on each of these model components.

\subsubsection{Complex-valued encoders}
First, we adapt complex-valued encoders to extract features from the complex-valued ultrasonic signals. This encoder is a complex-valued adaptation of the original RAFT~\cite{teed2020raft}. It takes input frames $I_1, I_2$ and applies an initial complex‑valued convolution that reduces the spatial resolution by a factor of two. This is followed by two residual blocks at half resolution, and then four residual blocks that maintain the same resolution. Consequently, the output feature maps $F_1, F_2$ have dimensions $F_1, F_2\in \mathbb{C}^{H_1\times W_1\times D}$ with $H_1 = H_o/2$ and $W_1 = W_o/2$, where $D=256$, is the number of complex-valued feature channels. The core operations in this encoder are complex-valued convolution ($\mathbb{C}\operatorname{Conv}$), complex-valued layer normalization ($\mathbb{C}\operatorname{LN}$), and complex-valued ReLU function ($\mathbb{C}\operatorname{ReLU}$). The complex-valued convolution ($\mathbb{C}\operatorname{Conv}$) uses complex-valued filters to capture both the magnitude and phase information of complex-valued inputs. For example, in the case of 2D convolution without bias, let $\mathbf{K}=(\Re(\mathbf{K})+ j\Im(\mathbf{K}) ) \in \mathbb{C}^{H_k\times W_k}$ represent a filter with size $H_k\times W_k$. The $\mathbb{C}\operatorname{Conv}$ operation on the complex-valued input $\mathbf{I_1}=(\Re(\mathbf{I_1})+ j\Im(\mathbf{I_1}) )$ can be expressed as 
% \begin{align\
\begin{equation}
    \begin{split}
        \mathbf{I_1} * \mathbf{K} &=(\Re(\mathbf{I_1})+j \Im(\mathbf{I_1})) *(\Re(\mathbf{K})+j \Im(\mathbf{K}))\\
&=\Re(\mathbf{I_1}) * \Re(\mathbf{K})-\Im(\mathbf{I_1}) * \Im(\mathbf{K})\\
&\quad\,\,\,\,+j(\Re(\mathbf{I_1}) * \Im(\mathbf{K})+\Im(\mathbf{I_1}) * \Re(\mathbf{K})).
    \end{split}
\end{equation}
The $\mathbb{C}\operatorname{LN}$ adopted in this work follows the formulation proposed in~\cite{2023building} to ensure that the model stabilizes forward propagation and mitigates the problem of vanishing or exploding gradients in back-propagation. Then, the complex nonlinear activation function is realized  by the operation $\mathbb{C}\operatorname{ReLU}(\cdot)$:
\begin{align}
\label{CReLU}
\mathbb{C} \operatorname{ReLU} \left(\cdot\right) =  \operatorname{ReLU}  \left(\Re(\cdot)\right) + j \operatorname{ReLU}  \left(\Im(\cdot)\right).
\end{align}
% \end{align}

\subsubsection{Correlation module}

Following feature extraction using complex-valued encoders, the correlation between feature maps $F_1$ and $F_2$ is computed to build a pixel-wise correlation volume. Conventional approaches implemented correlations of real-valued features via inner products; however, these operations are unsuitable for measuring similarity with complex-valued representations. Moreover, since the target estimated motion fields are real-valued, there exists an inherent need for both effective similarity measurement and domain conversion. To address this, we introduce a novel similarity measure that effectively captures inter-feature relationships within the complex-valued domain while simultaneously enabling transformation into a real-valued representation. This module operates as follows:
\begin{align}
\label{Crosscorrelation}
\mathcal{C} \left({F}_{1}, {F}_{2}\right)_{ijkl} &= \sum_{h}\Re \left({F}_{1, ijh} \cdot {F}_{2, klh}^{H}\right) \\ \nonumber
&=\sum_{h} |{F}_{1, ijh}||{F}_{2, klh}|cos\left(\theta_{1, ijh}-\theta_{2, klh}\right),
\end{align}
where $h$ indexes the feature channels (dimension $D$) of the complex‑valued feature maps ${F}_{1}$ and ${F}_{2}$, and the function $\mathcal{C} \left(\cdot, \cdot\right): \mathbb{C}^{H_1\times W_1 \times D} \times \mathbb{C}^{H_1\times W_1 \times D} \rightarrow \mathbb{R}^{H_1 \times W_1 \times H_1 \times W_1}$ constructs a real-valued 4D correlation volume from two complex-valued inputs. Each element within the volume reflects the cosine similarity between feature vectors at source positions $(i,j)$ and target coordinates $(k,l)$, weighted by their respective amplitudes. By integrating both phase difference and magnitude information, this formulation enables a smooth transition from complex to real-valued space, thereby supporting subsequent matching across frames for motion estimation. Then, a four‑level pyramid correlation volume is constructed from this base volume to capture motion at multiple scales, as in the original RAFT architecture~\cite{teed2020raft}. 

\subsubsection{A real-valued update operator}: The iterative refinement of the estimated motion field is governed by a real-valued recurrent update operator, inherited from the RAFT model~\cite{teed2020raft}. This operator employs a convolutional gated recurrent unit (GRU) to perform a series of updates. At each step $k$, the GRU receives the current motion estimate $\mathbf{f}_k$, relevant lookup values from the pyramid correlation volume, and the incoming hidden state. It then outputs an incremental update $\Delta\mathbf{f}$ to produce the next estimate $\mathbf{f}_{k+1} =  \mathbf{f}_k + \Delta \mathbf{f}$. By leveraging the context from previous steps through its hidden state and guided by the local correlation cues, this operator effectively converges on an accurate and fine-grained motion field through successive iterations. 

\subsection{Training Phase and Loss Function}
\subsubsection{Supervised learning}
 Our model was first trained on an open simulation phantom dataset~\cite{tehrani2020mpwc} via supervised learning. During training, we utilize the same loss function as proposed in~\cite{teed2020raft}. This function sums the $L_1$ loss iteratively with exponentially increasing weights as follows:
\begin{equation}\mathcal{L}_{sp}=\sum_{i=1}^T\gamma^{T-i}||\mathbf{f}_{gt}-\hat{\mathbf{f}}||_1,\end{equation}
where $\gamma$ is the weight of different iterations and set to 0.8, and $T$ is the total number of iterations.

\begin{table*}[ht!]
\centering
\caption{List of compared models and their properties. }
\resizebox{14cm}{!}{
% \begin{table}[]
\begin{tabular}{ccccc}
\hline
Method   & Input                  & Dimension of Input & Parameters & Inference time \\ \hline
MPWC++~\cite{tehrani2021mpwc++}    & RF + B-Mode + Envelope &  $\mathbb{R}^{H\times W\times 3}$                  & 3.56M       & 0.15s          \\
RAFT~\cite{teed2020raft}      & RF                     &  $\mathbb{R}^{H\times W\times 3}$                  & 5.26M     & 0.21s          \\
CrossRAFT & Analytic               &  $\mathbb{C}^{(H/4) \times (W/2) \times 3}$                  & 7.36M      & 0.65s          \\
CrossRAFT & IQ                     &  $\mathbb{C}^{(H/4) \times (W/2) \times 3}$                   & 7.36M      & 0.65s          \\ \hline
\end{tabular}}
% \end{table}
\label{Tab_benchmark}
\end{table*}

\subsubsection{Unsupervised learning}

\begin{figure}[ht!]
    
    \centering
    \includegraphics[scale=0.45]{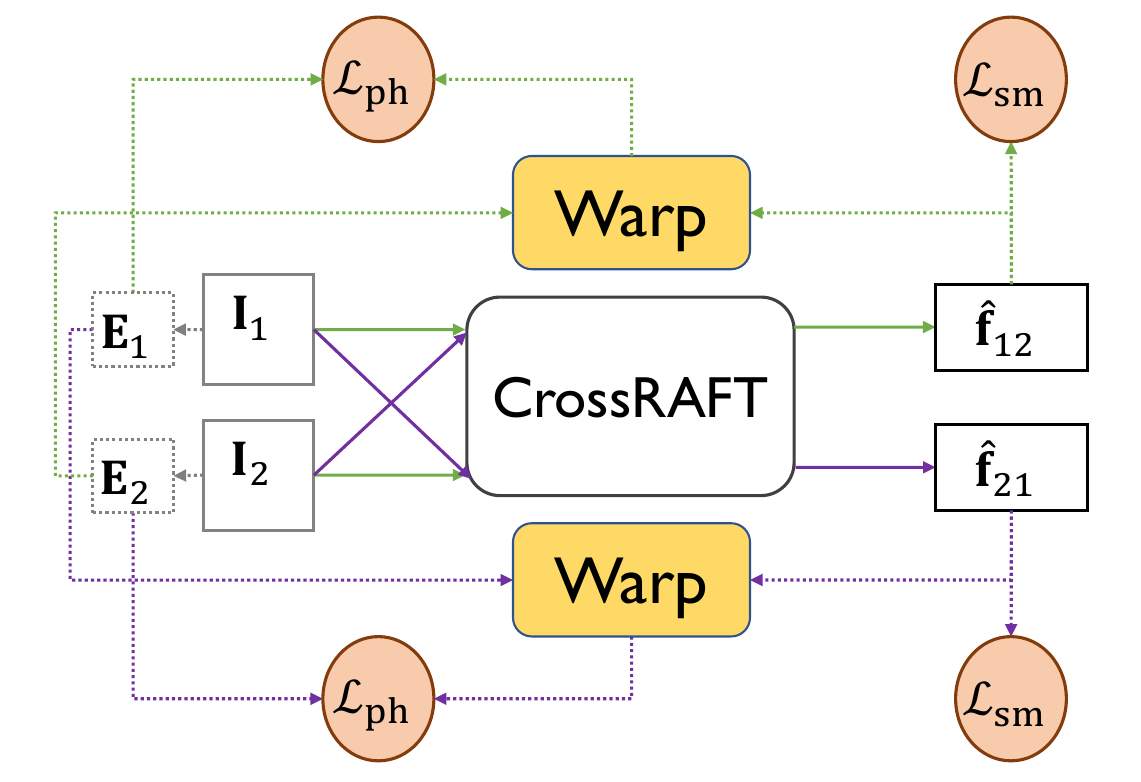}
    \caption{Unsupervised training pipeline: The model outputs forward and backward optical flows, indicated by green and purple arrows, respectively, with gradients propagated backward through two loss functions $\mathcal{L}_\text{ph}$ and $\mathcal{L}_\text{sm}$ jointly.}
    \label{Fig_unsp}
    \end{figure}

Clinical deployment of DL-based models for motion estimation requires robust generalization across heterogeneous \textit{in vivo} ultrasound data from diverse anatomical structures. However, GT motion annotations are inherently unobtainable from clinical data. To address this domain shift, we adapt our model to a simulated echo dataset~\cite{burman2024COLE} and private \textit{in vivo} ultrafast echocardiographic data~\cite{xu2025near} in an unsupervised learning way. In practice, we instantiate the generic similarity measure $\mathcal{S}$ as a photometric consistency loss $\mathcal{L}_\text{ph}$, and the regularization term $\mathcal{R}$ as a discontinuity‑aware smoothness constraint $\mathcal{L}_\text{sm}$. The overall unsupervised loss is then formulated as:
\begin{equation}
    \mathcal{L}_\text{unsp} = \mathcal{L}_\text{ph} + \lambda \mathcal{L}_\text{sm}, 
\end{equation}
 where $\lambda=2.0$ retains the same role of balancing the similarity and smoothness terms. Instead of operating directly on RF, analytic, or IQ signals, our unsupervised loss is computed on the envelope data $(\mathbf{E}_1, \mathbf{E}_2)$, which retains the anatomical structure information while suppressing the rapid phase oscillations and speckle noise that can mislead the photometric consistency objective. This design encourages the model to focus on tissue motion patterns. As illustrated in Fig.~\ref{Fig_unsp}, the model generates bidirectional optical flow fields optimized jointly via $\mathcal{L}_\text{ph}$ and $\mathcal{L}_\text{sm}$.

The photometric consistency loss $\mathcal{L}_\text{ph}$ is formulated using the SSIM to enforce the consistency of intensity appearance between the warped target frame and the source frame $\mathbf{I}_{t}$:
 \begin{equation}
    \mathcal{L}_\text{ph} = \frac{1}{P}\sum_{i=1}^{P}\left(1- \operatorname{SSIM}(\mathbf{E}_{t+1}(\mathbf{x}_i+\hat{\mathbf{f}}_{12}(\mathbf{x}_i)), \mathbf{E}_{t}(\mathbf{x}_i))\right),
\end{equation}
where $P$ is the total number of pixels in $\mathbf{E_t}$, $\mathbf{x}$ denotes pixel coordinates, and $\hat{\mathbf{f}}_{12}$ represents forward displacement vectors. The function $\operatorname{SSIM}\left(\cdot, \cdot \right)$ evaluates local structural similarity over a $3 \times3$ Gaussian‑weighted window centered at $\mathbf{x}_i$. For bidirectional consistency, this loss is applied to the reverse flow $\hat{\mathbf{f}}_{21}$. 

The smoothness constraint $\mathcal{L}_\text{sm}$ is an edge-preserving smoothness loss developed by~\cite{heise2013smloss} that regularizes flow gradients while preserving motion boundaries through anisotropic weighting:
 \begin{equation}
    \mathcal{L}_\text{sm} = \sum_{d \in \{x,y\}}\frac{1}{P}\sum_{i=1}^P\Vert \nabla_d \mathbf{f}_{12} (\mathbf{x}_i) \Vert_1 \odot e^{-\beta \Vert\nabla_d \mathbf{G_t}(\mathbf{x}_i) \Vert_1},
\end{equation}
where $\nabla_d$ computes the spatial (i.e., axial and lateral) gradients of $\mathbf{G_t}$, $\beta$, set to 50, controls the sensitivity to image edges, and $\odot$ denotes the Hadamard product for per‑pixel gradient weighting. As with the photometric loss, this smoothness term is also applied to the reverse flow $\hat{\mathbf{f}}_{21}$. 

\subsection{Performance Evaluation}
 The proposed CrossRAFT framework was evaluated against a conventional NCC-based method~\cite{li2016systematic} and two established DL benchmarks: MPWC++~\cite{tehrani2021mpwc++} and the original RAFT model~\cite{teed2020raft}. The details of these DL benchmark models are summarized in Table~\ref{Tab_benchmark}. All inference times were measured on the test set of the simulated phantom dataset, averaged over one frame pair, using a single NVIDIA GeForce RTX 4090 GPU. Note that inference times may vary across different datasets due to differences in image size or input preprocessing. 
 
 MPWC++~\cite{tehrani2021mpwc++} is a modified framework of PWC-Net~\cite{sun2018pwc}, which employs a multichannel input strategy by concatenating beamformed RF signals, B-mode images, and envelope data. This model follows a two-stage training regimen. It is first pre-trained on large-scale optical flow datasets (e.g., FlyingChairs~\cite{FlyingChairs2015flownet}) and subsequently fine-tuned in an unsupervised way on a simulated phantom dataset. Since MPWC++~\cite{tehrani2021mpwc++} was not used in the cardiac scenario, it is excluded from the cardiac dataset experiments. The original RAFT architecture~\cite{teed2020raft} is applied to beamformed RF signals with pre-training on the simulated phantom dataset~\cite{tehrani2020mpwc}.
    
Unlike the original RAFT that processes real-valued RF data, the proposed CrossRAFT operates on lower-dimensional, complex-valued analytic or IQ signals. To ensure a fair comparison despite different input dimensions, the encoder convolutional downsampling rates were carefully adjusted in each comparison network so that the final feature resolution at the bottleneck was identical across RAFT and CrossRAFT. This controlled design isolates the impact of the input representation and complex-valued feature learning from any differences that arise from spatial resolution. 
  
The performance of the proposed CrossRAFT was evaluated by three metrics. On the phantom dataset, where ground‑truth strains were available across different strain levels, we used the Normalized Root Mean Squared Error (NRMSE) to measure the accuracy of both displacement and strain estimation, and the elastographic Contrast‑to‑Noise Ratio (CNR$_e$)~\cite{bilgen1997predicting} to quantify the quality of the strain maps. On the cardiac dataset, Mean Absolute Error (MAE) was used for displacement accuracy. Specifically, NRMSE normalizes the root‑mean‑square error by the maximum value of the ground‑truth, providing a scale‑independent measure of estimation accuracy:
  \begin{equation}
    \text{NRMSE} = \frac{
        \sqrt{ \frac{1}{P} \sum_{p=1}^{P} \left( [\mathbf{f}_{\text{GT}}]_p - [\mathbf{\hat{f}}]_p \right)^2 }
    }{
        \max_{p} \left| [\mathbf{f}_{\text{GT}}]_p \right|
    } \times 100\%,
\end{equation}
    where $p$ indexes the $P$ pixels in the displacement field, $\hat{\mathbf{f}}$ represents the estimated 2D displacement field and $\mathbf{f}_\text{GT}$ is the ground truth motion field. Lower NRMSE values indicate higher estimation accuracy.
    The CNR$_e$ quantifies the contrast between the target region and the background in the strain map, indicating how distinctly the inclusion can be distinguished from the surrounding tissue, which is given by:
    \begin{equation}
    \text{CNR}_e=\sqrt{\frac{2(\mu_b-\mu_t)^2}{\sigma_b^2+\sigma_t^2}},
\end{equation}
where $\mu_b$ and $\mu_t$ denote mean strain values in background regions and ROI region (e.g., lesion), respectively, $\sigma_b^2$ and $\sigma_t^2$ represent corresponding standard deviations. Higher CNR$_e$ values reflect improved clinical interpretability of tissue inhomogeneity. MAE compares the estimated pixel-wise displacement vector
to the GT as follows: 
 \begin{equation}
    \text{MAE}=\frac{1}{P}\parallel\mathbf{f}_\text{GT}-\mathbf{\hat{f}}\parallel_1.
\end{equation}
Lower MAE values indicate higher estimation accuracy.
\section{Datasets and Experiments}
\label{sec_3}

\begin{figure}[ht!]
    \centering
    \includegraphics[scale=0.35]{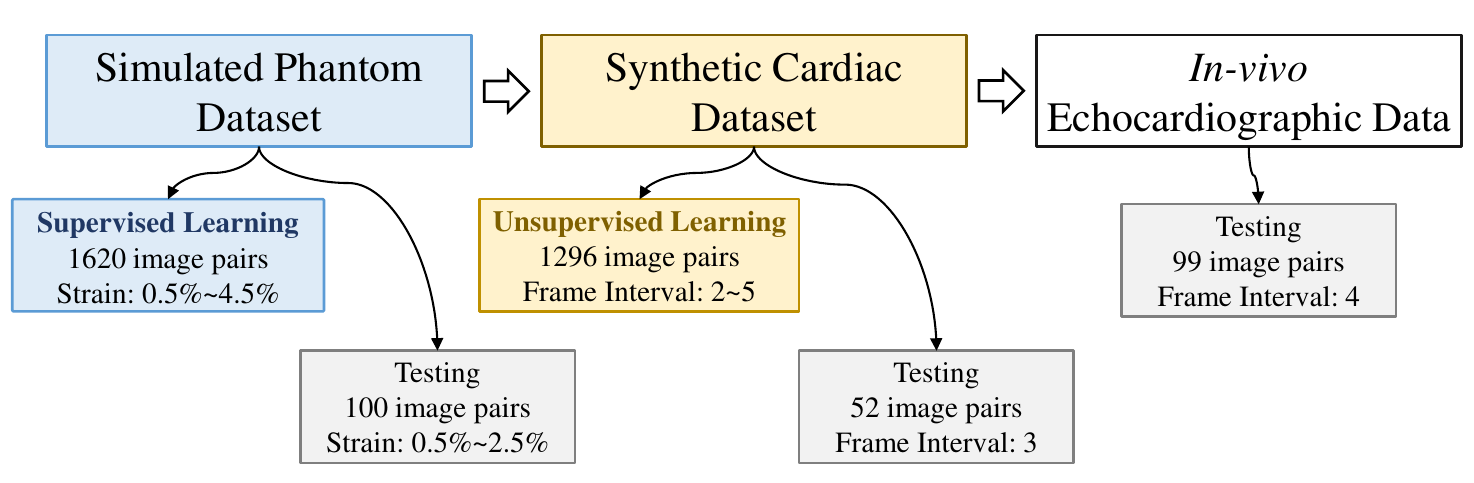}
    \caption{Workflow and configuration of the datasets used in this work.}
    \label{Fig_dataset}
    \end{figure}
The overall dataset workflow is illustrated in Fig.~\ref{Fig_dataset}. The proposed CrossRAFT model was trained on the simulated phantom~\cite{tehrani2020mpwc} dataset in a supervised manner, and then adapted to the synthetic cardiac ultrasound dataset~\cite{burman2024COLE} via unsupervised learning. \textit{In vivo} echocardiographic data~\cite{xu2025near} were used for clinical validation. The details of each dataset are described below.
\subsection{Simulated Phantom Data}
The simulated phantom dataset~\cite{tehrani2020mpwc} has a total of 24 distinct phantoms, each of which contains one or two inclusions with random positions, covering ten different average applied strain levels and ten independent random scatterer distributions. The applied strain level varied from $0.5\%$ to $4.5\%$, and the displacement fields were obtained from finite element (FE) simulations performed in ABAQUS. In the supervised training phase, a total of 1620 image pairs were randomly selected to ensure diversity across strain levels and scatterer realizations. For testing, 100 image pairs were used, covering five different scatterer realizations and average strain levels between $0.5\%$ and $2.5\%$.
 
\subsection{Synthetic Cardiac Ultrasound Data}
\begin{figure}[ht!]
    \centering
    \includegraphics[scale=0.42]{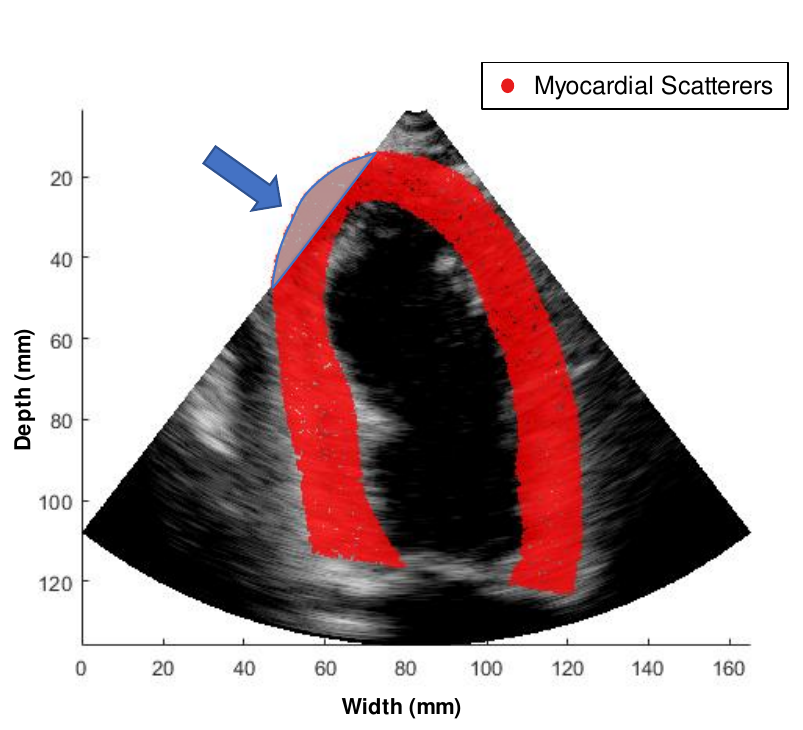}
    \caption{Illustration of one B‑mode frame from~\cite{burman2024COLE} with overlaid myocardial scatterers (red points), where scatterers inside the blue solid line are considered out‑of‑view and excluded from computation.}
    \label{Fig_cole}
    \end{figure}
This work uses a synthetic cardiac ultrasound dataset~\cite{burman2024COLE} comprising 1296 diverse and realistic simulated ultrasound recordings in both two‑chamber (2-CH) and four‑chamber (4-CH) apical views. Each recording provides ground‑truth left ventricular (LV) motion and geometry, beamformed RF signals, and scatterer maps (scatterer positions and amplitudes). In this dataset, the speckle appearance is derived from real clinical scans, whereas the underlying LV deformation is generated using the CircAdapt heart model. Fig.~\ref{Fig_cole} illustrates one B‑mode frame from the recording of Patient 173, and red points denote scatterers located within the heart wall. Inter-frame LV wall motion is computed from the coordinate displacements of these scatterers. Scatterers located inside the region enclosed by blue solid curves move out of the imaging plane during the cardiac cycle and are therefore excluded from motion estimation.

In this work, we selected 12 recordings from Patient 108 to fine-tune and 1 recording from Patient 173 to evaluate performance. Each recording usually has 50-55 frames. The frame interval between pre-deformed and post-deformed data was varied from 2 to 5 frames to increase motion diversity for training. A fixed interval of 3 frames was used for consistent comparison at evaluation. Notably, the input to the DL models (RAFT and CrossRAFT) is represented in Cartesian coordinates after scan conversion, whereas the conventional NCC method operates on polar‑coordinated RF signals and its resulting displacement estimates are subsequently scan‑converted to Cartesian space for fair comparison with the DL‑based methods.

\subsection{\textit{In vivo} Echocardiographic data}
To further validate the proposed CrossRAFT framework under realistic clinical conditions, we included \textit{in vivo} echocardiographic data acquired from a healthy subject. The data were obtained using a Vantage 256 system equipped with a P4-2 probe with a center frequency of 2.5 MHz. B-mode images of the LV were acquired from the apical 2-CH view. The frame rate was maintained at 400 fps to ensure adequate temporal resolution for capturing rapid cardiac motion throughout the entire cardiac cycle. The study was approved by the Institutional Review Board of The University of Hong Kong (ref.: UW 19-043), and written informed consent was obtained from the participant. A fixed interval of 4 frames was used during inference. The input coordinate settings follow the same protocol as for the synthetic cardiac dataset. 

\section{Results and Analysis}
\label{sec_result}
\subsection{Simulated Phantom Dataset}

\begin{figure*}[htbp]
    \centering
    \begin{minipage}[b]{0.75\textwidth}
        \centering
        \includegraphics[width=\linewidth]{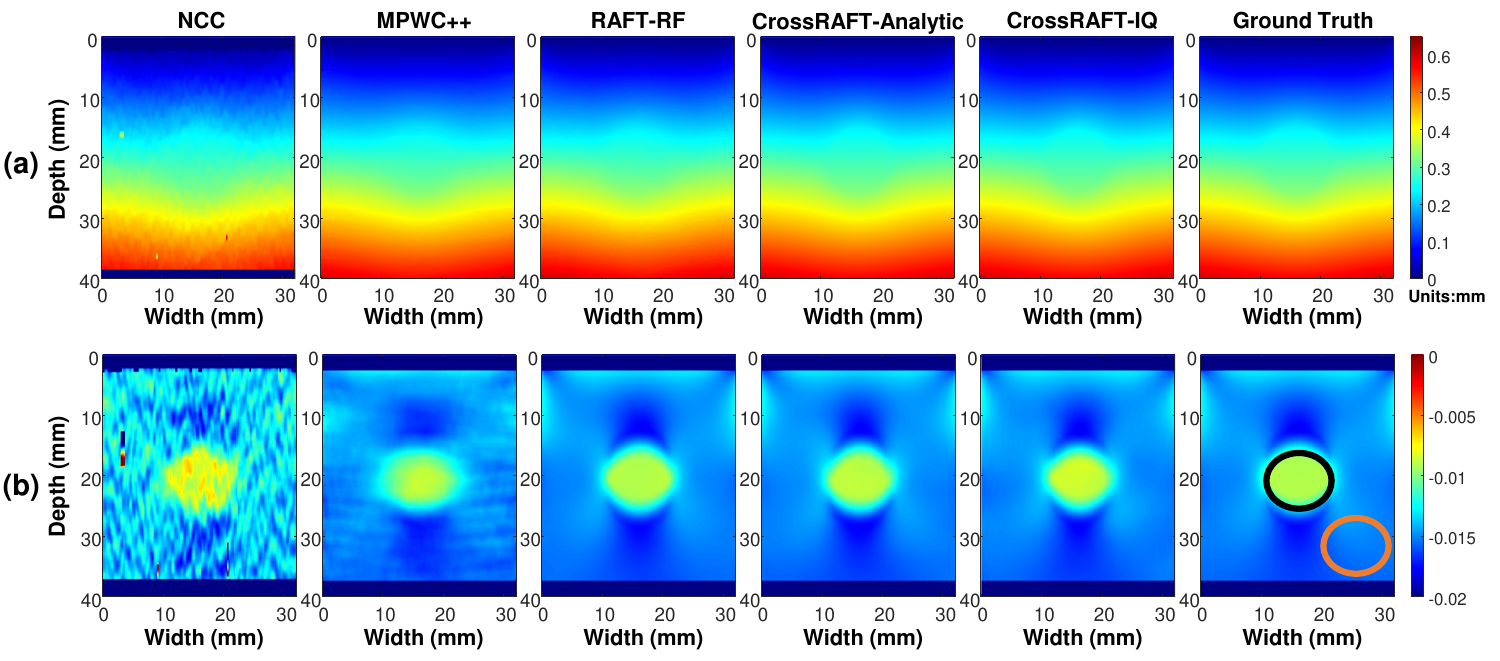}
        % \label{fig:sub1}
    \end{minipage}
    \begin{minipage}[b]{0.75\textwidth}
        \centering
        \includegraphics[width=\linewidth]{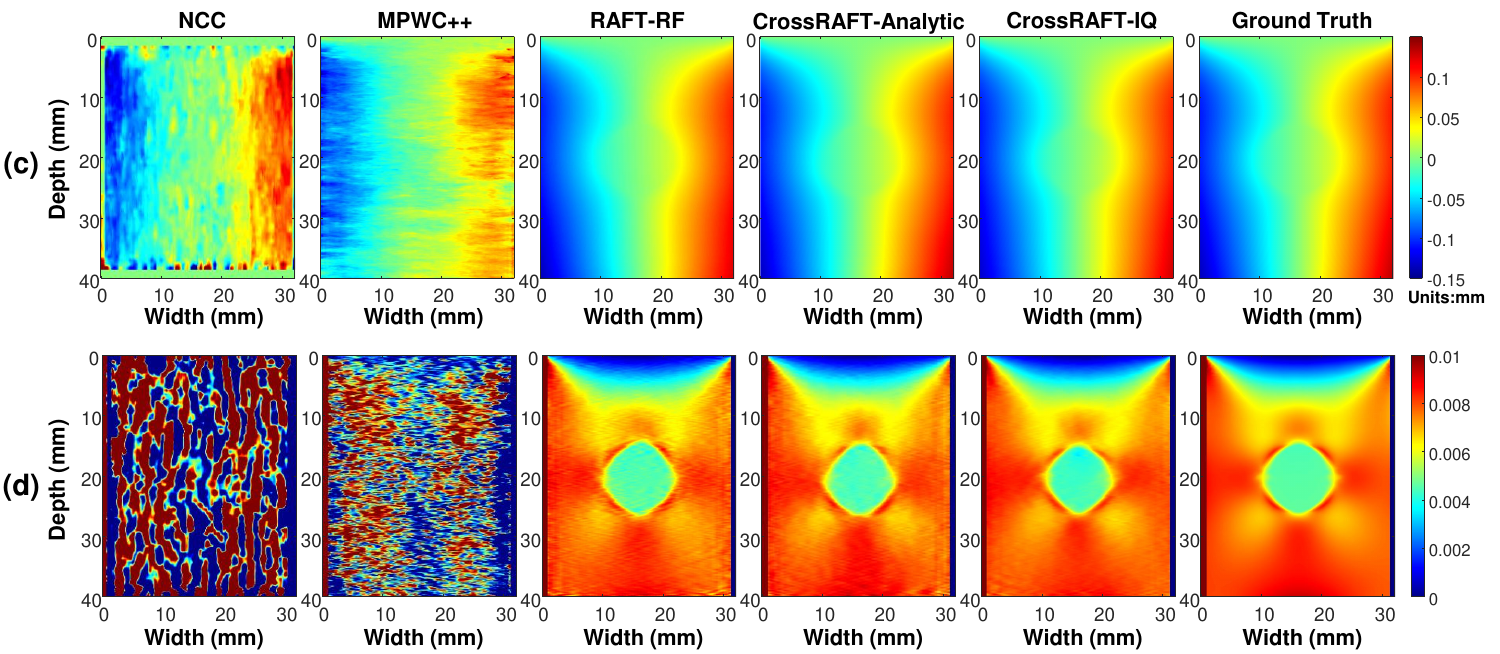}
        % \label{fig:sub2}
    \end{minipage}
    \caption{Results on simulated phantom dataset: \textbf{(a)} Axial displacement (all positive values: upward motion), \textbf{(b) }axial strain (all negative values: axial compression), \textbf{(c)} lateral displacement (positive value: rightward motion; negative value: leftward motion), and \textbf{(d)} lateral strain (all positive values: lateral extension) images obtained by one traditional method (NCC), four DL-based models (MPWC++, RAFT, and our proposed CrossRAFT for analytic and IQ data, respectively), and the ground truth (rightmost column). The inclusion (black circle) and background ROI (orange circle) in the ground truth axial strain image in (b) were used to calculate CNR$_e$}
    \label{Fig_phantom}
\end{figure*}

\begin{table*}[t!]
\caption{Normalized root mean square errors (NRMSEs) of displacements and contrast-to-noise ratio (CNR) of strains (mean $~\pm~$deviation) across the $0.5\%-2.5\%$ strain cases in the phantom dataset.}
\resizebox{\textwidth}{13mm}{
\begin{tabular}{ccccccc}
\hline
\multirow{2}{*}{Method} & \multicolumn{2}{c}{Displacement Estimation}  & \multicolumn{4}{c}{Strain Map}                                                \\ 
                        & NRMSE of Axial ($\%$)   & NRMSE of Lateral ($\%$) &NRMSE of Axial ($\%$) & NRMSE of Lateral ($\%$) & CNR$_e$ in Axial & CNR$_e$ in Lateral \\ \hline
NCC~\cite{li2016systematic}                   & 14.6$~\pm~$15.8            & 37.4$~\pm~$14.4             & 22.7$~\pm~$15.3         & 4.70$~\pm~$3.56            & 6.13$~\pm~$9.45 & -7.73$~\pm~$4.77   \\

MPWC++~\cite{tehrani2021mpwc++}                   & 0.72$~\pm~$0.35            & 16.7$~\pm~$2.91            & 0.231$~\pm~$0.062           & 0.480$~\pm~$0.164              & 21.3$~\pm~$2.52 & -18.4$~\pm~$10.1  \\
RAFT-RF                     & \textbf{0.10$~\pm~$0.13}           & 1.55$~\pm~$1.75            & \textbf{0.039$~\pm~$0.022 }           & 0.025$~\pm~$0.021              & \textbf{25.2$~\pm~$1.91}   & \textbf{21.4}$~\pm~$4.72     \\
CrossRAFT-Analytic             & 0.17$~\pm~$0.16           & \textbf{1.40$~\pm~$0.82}             & 0.075$~\pm~$0.018           & \textbf{0.024}$~\pm~$\textbf{0.010}              & 23.0$~\pm~$2.80   & 21.2$~\pm~$3.34    \\ 
CrossRAFT-IQ             & 0.21$~\pm~$0.16           & 1.69$~\pm~$1.41             & 0.117$~\pm~$0.047            & 0.030$~\pm~$0.024             & 21.6$~\pm~$1.96   & 19.6~$\pm$~\textbf{2.59}     \\  \hline 
\end{tabular}
} 
\label{Tab_phantom}
\end{table*}
Fig.~\ref{Fig_phantom} shows 2D (i.e., axial and lateral) displacement and strain maps obtained by different methods against the GT at 1.5\% axial compression of the simulated phantom dataset. The displacement maps estimated by RAFT and CrossRAFT are closer to the GT than those of NCC and MPWC++ in both axial (Fig. ~\ref{Fig_phantom}(a)) and lateral (Fig. ~\ref{Fig_phantom}(c)) directions. Their corresponding strain maps (Fig. ~\ref{Fig_phantom}(b) and (d)) also show excellent quality, where the inclusion is well differentiated from the background. This is especially evident in the lateral direction, in contrast to the noisy strain maps produced by MPWC++ and NCC. Table~\ref{Tab_phantom} summarizes the average displacement NRMSEs, strain NRMSEs, and CNR$_{e}$ values across the $0.5\%-2.5\%$ average strain cases. Among all methods, RAFT-RF yields the lowest displacement and strain NRMSEs in the axial direction. However, in the lateral direction, the proposed CrossRAFT framework with analytic inputs produces the lowest lateral displacement NRMSE and strain NRMSE, whereas maintaining CNR$_{e}$ values in both axial and lateral directions competitive with those of RAFT‑RF.In addition, CrossRAFT-IQ achieves displacement and strain estimation accuracies comparable to or slightly lower than RAFT‑RF.

\subsection{Synthetic Cardiac Ultrasound Data}
\begin{figure*}[ht!] 
	\centering
	\subfigure[]{ 
		\label{Fig_19}
		\includegraphics[width=0.65\textwidth]{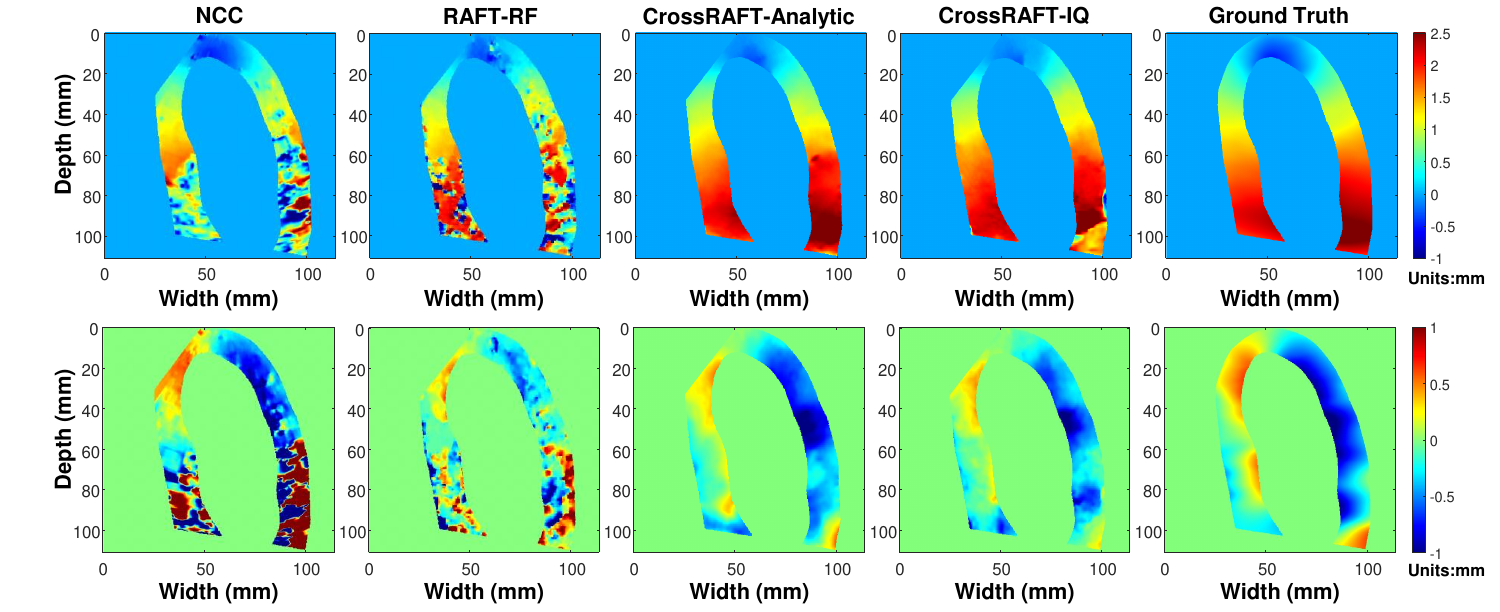}} \hspace{0in}
        \subfigure[]{ 
		\label{Fig_41}
		\includegraphics[width=0.65\textwidth]{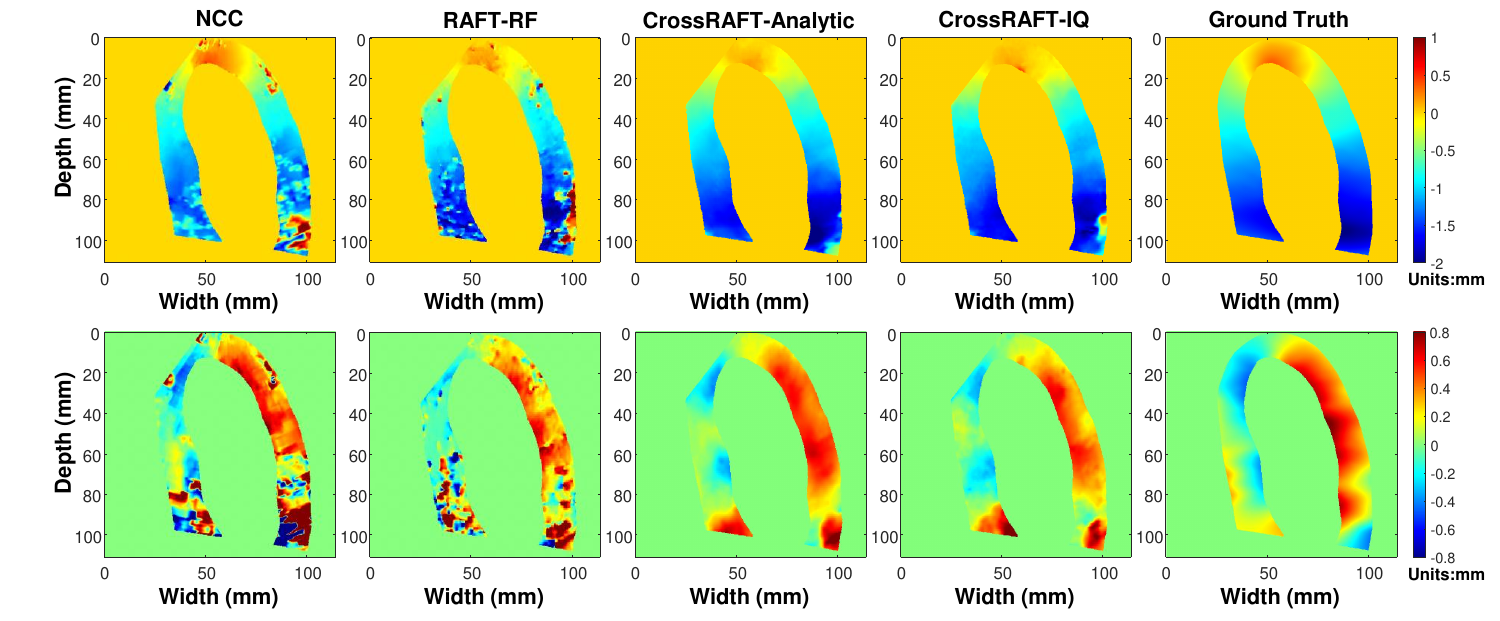}} \hspace{0in}
	\caption{Results on a synthetic \textit{cardiac} dataset: Estimated displacement maps in vertical and horizontal directions across the evaluated methods at systolic \textbf{(a)} and diastolic \textbf{(b)} phases. The top rows of \textbf{(a)} and \textbf{(b)} represent vertical displacements (all positive values: upward motion), and the bottom rows are horizontal displacements (positive value: rightward motion; negative value: leftward motion). A supplementary video is provided to show a side-by-side comparison of the estimated 2D displacement maps across all methods over a complete cardiac cycle.}  \label{Fig_cardiac}
\end{figure*}

Fig.~\ref{Fig_cardiac} shows 2D displacement maps of the human heart in the apical 2-CH view during the systolic and diastolic phases. The vertical displacement maps in Fig.~\ref{Fig_19} show an upward LV wall motion during systole. Concurrently, the LV posterior wall exhibits a rightward motion, and the LV anterior wall moves leftward. A similar but opposite motion field is observed in Fig.~\ref{Fig_41} during diastole. These results are consistent with the established cardiac biomechanics for the normal heart.

As shown in Table~\ref{Tab_cardiac}, all evaluated methods provide reasonable vertical displacement estimates. Among them, CrossRAFT (both analytic and IQ variants) achieves higher accuracy than RAFT‑RF and NCC in both vertical and horizontal displacement estimation, confirming the advantage of complex‑valued signal processing. Notably, Fig.~\ref{Fig_cardiac} shows that both NCC and RAFT‑RF deliver unsatisfactory performance with noise.

\begin{table}[t!]
\caption{Mean absolute errors of displacements (MAEs) (mean $~\pm~$deviation) in the synthetic cardiac dataset.}
\label{Tab_cardiac}
\resizebox{\linewidth}{10mm}{
\begin{tabular}{ccc}
\hline
\multirow{2}{*}{Method} & \multicolumn{2}{c}{Displacement Estimation} \\ \cline{2-3} 
                        & MAE of Vertical (mm)   & MAE of Horiztonal (mm)   \\ \hline
NCC~\cite{li2016systematic}                     & $0.309 \pm 0.071$              & $0.398\pm 0.051$                \\
RAFT-RF                    & $0.266 \pm 0.049$             & $0.252 \pm 0.022$                \\
CrossRAFT-Analytic      & $\textbf{0.082}\pm \textbf{0.011}$            & $0.123 \pm \textbf{0.010}$                \\
CrossRAFT-IQ           & $0.104 \pm 0.016$               & $\textbf{0.122} \pm 0.014$                \\ \hline
\end{tabular}}
\end{table}
\subsection{\textit{In vivo} Echocardiographic Data}
\begin{figure*}[ht!] 
	\centering
	\subfigure[]{ 
		\label{Fig_MRP32}
		\includegraphics[width=0.55\textwidth]{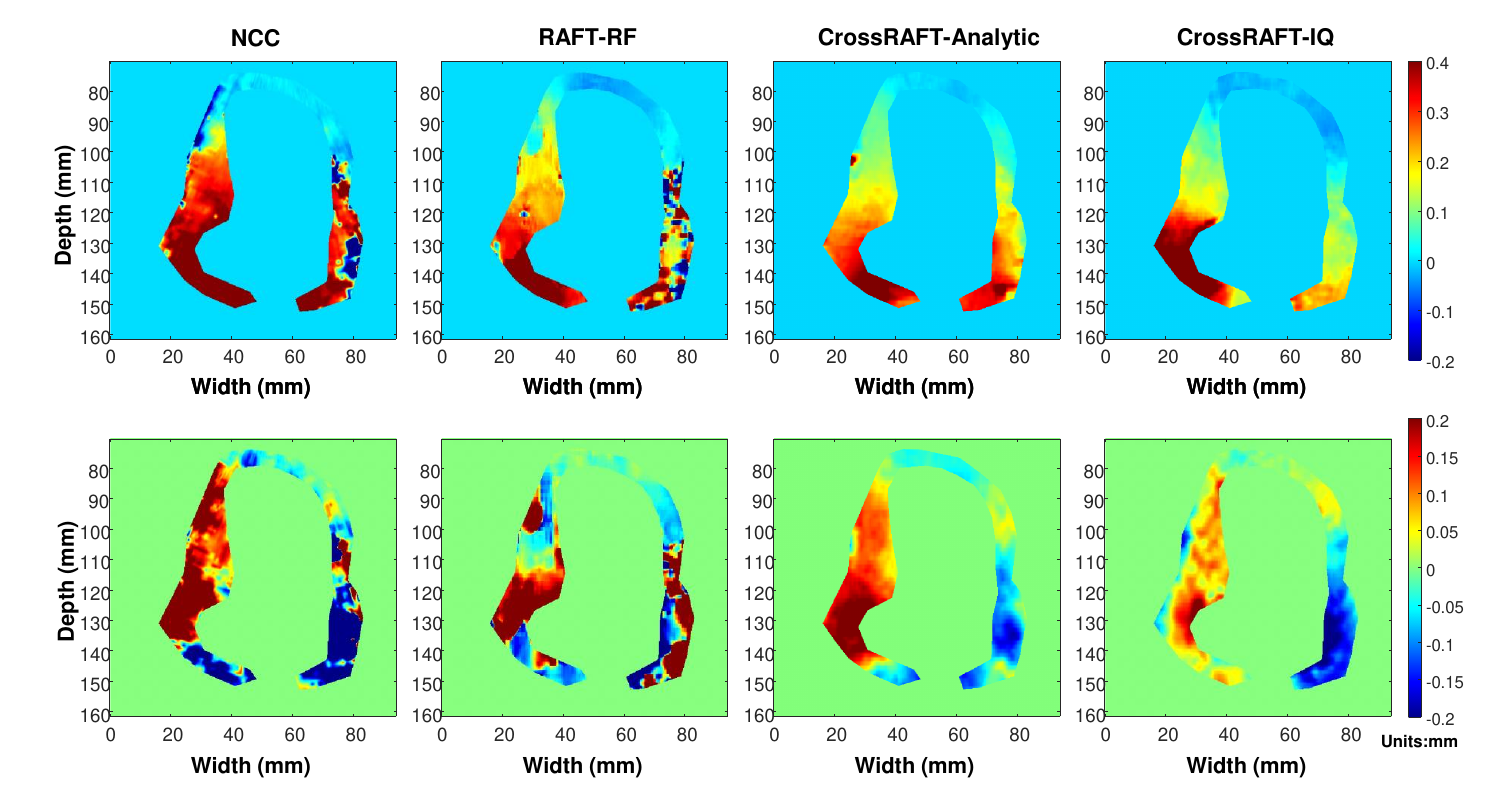}} \hspace{0in}
        \subfigure[]{ 
		\label{Fig_MRP69}
		\includegraphics[width=0.55\textwidth]{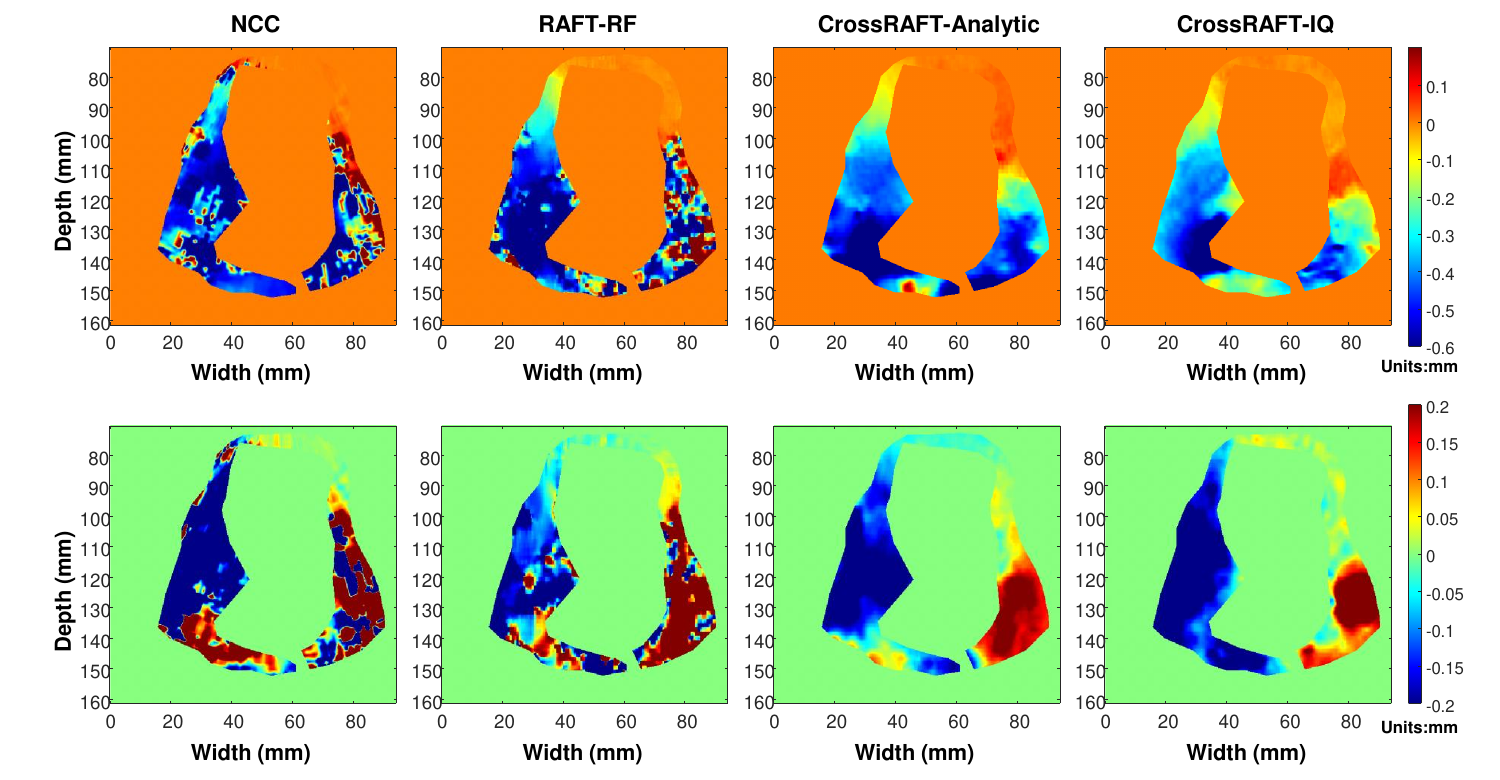}} \hspace{0in}
	\caption{The experimental results on \textit{in vivo} data: Comparison of estimated displacement maps in vertical and horizontal directions across the evaluated methods at systolic \textbf{(a)} and diastolic \textbf{(b)} phases. The top rows of \textbf{(a)} and \textbf{(b)} represent vertical displacements (all positive values: upward motion), and the bottom rows are horizontal displacements (positive value: rightward motion; negative value: leftward motion). A supplementary video is provided to show a side-by-side comparison of the estimated 2D displacement maps across all methods over a complete cardiac cycle.}  \label{Fig_invivo}
\end{figure*}

Fig.~\ref{Fig_invivo} shows the displacement fields estimated by the proposed CrossRAFT (analytic and IQ data), RAFT-RF, and NCC during systole (Fig.~\ref{Fig_MRP32}) and diastole (Fig.~\ref{Fig_MRP69}).

During systole (Fig.~\ref{Fig_MRP32}), upward LV wall motion (top row) alongside rightward posterior wall motion and leftward anterior wall motion is consistent with known cardiac mechanics of the normal heart. CrossRAFT produces displacement fields that closely matches the expected mechanical pattern, with less estimation noise compared to RAFT-RF and NCC~\cite{li2007quantification}. Similar performance results are observed during diastole (Fig.~\ref{Fig_MRP69}), where the left ventricle exhibits opposite motion patterns. 

\subsection{Ablation Study}
\begin{figure}[ht!]
    \centering
    \includegraphics[scale=0.42]{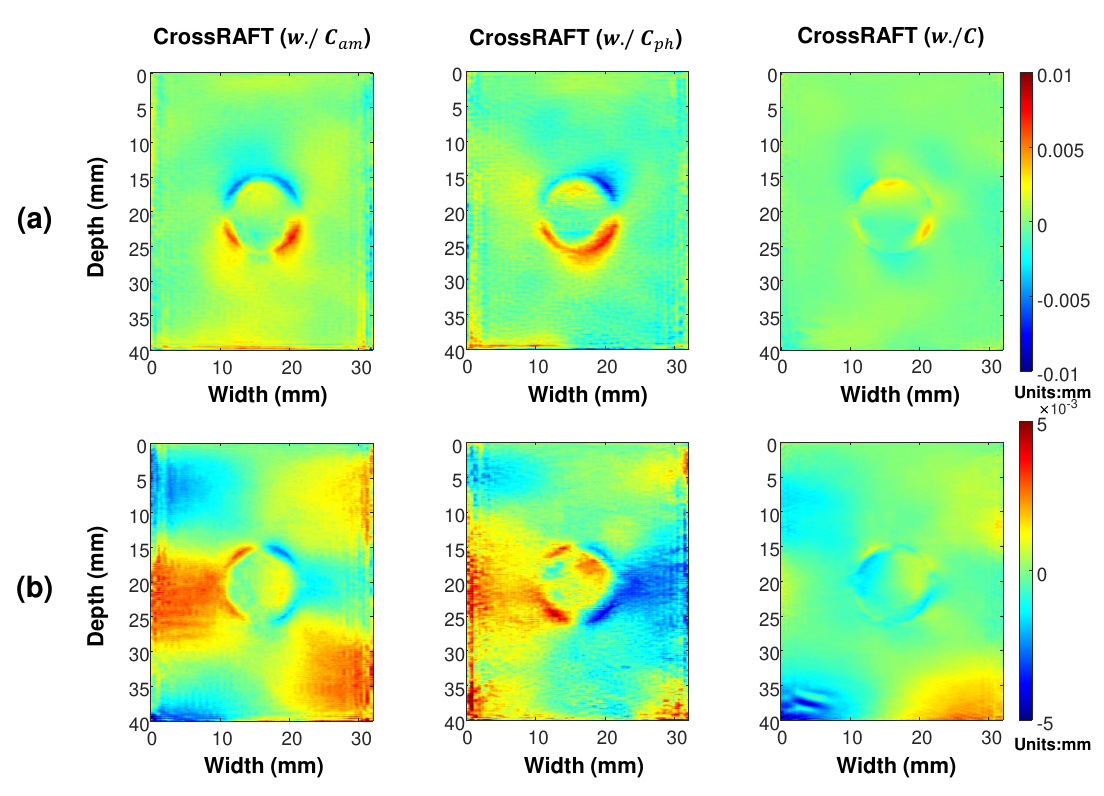}
    \caption{Difference maps between the displacement fields estimated by three CrossRAFT variants with different correlation modules (the proposed $\mathcal{C}$, amplitude‑only $\mathcal{C}_{am}$, and phase‑only $\mathcal{C}_{ph}$) and the ground truth on \textbf{(a)} axial and \textbf{(b)} lateral directions.}
    \label{Fig_ab}
    \end{figure}
To validate the design of our proposed correlation module, we conduct an ablation study on the phantom dataset, comparing its performance against variants that use either only amplitude or only phase information:
\begin{itemize}

    \item Amplitude-only Correlation: \begin{equation}
        \mathcal{C}_{am} \left({F}_{1}, {F}_{2}\right)_{ijkl} =\sum_{h} |{F}_{1, ijh}|\cdot|{F}_{2, klh}|
        \end{equation}
    \item Phase-only Correlation: \begin{equation}
        \mathcal{C}_{ph} \left({F}_{1}, {F}_{2}\right)_{ijkl} =\sum_{h} cos\left(\theta_{1, ijh}-\theta_{2, klh}\right)
    \end{equation}

\end{itemize}

Table~\ref{Tab_ablation} shows that the full model achieves the lowest displacement NRMSEs.  Using only phase information leads to performance degradation, increasing axial and lateral errors by 157\% and 45.0\%, respectively. Similarly, relying solely on amplitude information increases the axial MAE by 71.4\% and lateral NRMSE by 40.8\%. Compared to the phase‑only variant,  the amplitude‑only variant provides a statistically significant improvement in axial displacement estimation (paired t‑test, $p < 0.001$), whereas the difference in the lateral direction is not statistically significant ($p=0.3385$).

Fig.~\ref{Fig_ab} visualizes the difference maps between the estimates of the three CrossRAFT variants and the ground truth. For both the amplitude‑only $\mathcal{C}_{am}$ and phase‑only ($\mathcal{C}_{ph}$) variants, larger errors are observed near the inclusion boundaries. These results indicate that both phase and amplitude components are essential for accurate motion estimation by preserving sharp boundaries at inclusion edges.

\begin{table}[]
\caption{Ablation study of the proposed correlation module}
\resizebox{\linewidth}{8mm}{
\begin{tabular}{ccc}
\hline
\multirow{2}{*}{Model Variant}        & \multicolumn{2}{c}{Displacement Estimation} \\
                                      & NRMSE of Axial ($\%$)     & NRMSE of Lateral ($\%$) \\ \hline
CrossRAFT (w/. $\mathcal{C}$)                           & 0.21$~\pm~$0.16~(--)         & 1.69$~\pm~$1.41~(--)                \\
CrossRAFT (w/. $\mathcal{C}_{am}$)  & $0.36~\pm~0.13~(\uparrow 71.4\%)$  & $2.38~\pm~1.12 ~(\uparrow40.8\%)$ \\
CrossRAFT (w/. $\mathcal{C}_{ph}$)    & $0.54~\pm~0.36~(\uparrow157\%)$   & $2.45~\pm~1.27 ~(\uparrow45.0\%)$ \\
 \hline
\end{tabular}}
\label{Tab_ablation}
\end{table}
\section{Discussion}
\label{sec_dis}
CrossRAFT employs complex‑valued feature encoders and a custom correlation module for motion estimation directly from analytic ultrasound signals. Quantitative results of simulated phantom, synthetic echocardiographic, and private \textit{in vivo} echocardiographic data demonstrate how preserving phase information offers an effective alternative solution for motion estimation in ultrasound imaging.

On the simple phantom with a uniform background and stiff inclusions (Table~\ref{Tab_phantom}),  CrossRAFT-Analytic produces better displacement estimation and strain maps in lateral direction than RAFT‑RF. This demonstrates that preserving phase information through complex‑valued processing is particularly beneficial for lateral motion estimation. The benefit of complex-valued inputs becomes more evident in Table~\ref{Tab_cardiac} on the synthetic cardiac dataset, where tissue motion is multi‑directional and rapid. On this dataset, CrossRAFT with analytic signal inputs yielded the lowest displacement errors in both directions, whereas CrossRAFT‑IQ also performed nearly as well, with only a slightly higher axial MAE. 

These results underscore that phase information is indispensable for accurate motion estimation. Although IQ data are band‑pass filtered and susceptible to attenuation‑induced information loss, they still confer substantial phase benefits over real‑valued RF inputs. The analytic signal offers a mathematically favorable representation because it naturally preserves instantaneous amplitude and phase, directly supplying the network with the features most relevant for displacement estimation. 

The supervised RAFT‑RF model achieved the lowest axial displacement errors on the simulated phantom dataset as shown in Table~\ref{Tab_phantom}, demonstrating its ability to accurately capture motion from real‑valued RF inputs to displacement fields when the motion scenario is relatively simple. However, on the synthetic cardiac dataset, RAFT‑RF exhibited notably higher errors than CrossRAFT (Fig. \ref{Fig_cardiac} and Table \ref{Tab_cardiac}). This can be attributed to two main factors. Firstly, the cardiac data were generated with realistic speckle texture derived from clinical scans, and contain rapid and multi‑directional motion throughout the cardiac cycle. The supervised RAFT‑RF, trained exclusively on phantom data with quasi-static compression, did not generalize to these more complex motion patterns well, likely due to overfitting to the specific texture and deformation statistics of the phantom. Secondly, the cardiac dataset benefits more from explicit phase information because phase variations in complex‑valued ultrasound signals encode tissue motion and remain reliable even when amplitude decorrelates severely~\cite{yuan2015analytical}. Therefore, in the rapid, multi-directional cardiac motion scenario, the underlying phase information remains stable for reliable motion estimation. Real‑valued RF signals lack an explicit phase representation, which forces the network to implicitly learn phase‑like features. In contrast, CrossRAFT directly processes the analytic signal, enabling the network to exploit the continuous phase domain for precise motion estimation despite complex multi-directional deformations. This interpretation aligns with the findings of Vinals \textit{et al.}~\cite{vinals2024IQvsRF}, who compared training CNNs with RF versus IQ images to enhance single‑plane‑wave imaging. Specifically, whereas training with RF yielded higher PSNR and SSIM, IQ-based training better preserved speckle pattern resolution, which is crucial for accurate speckle tracking.

A closely related work to our CrossRAFT is RAFT‑USENet~\cite{majumder2025useraft}. Both methods build upon the RAFT architecture and target ultrasound motion estimation. However, RAFT‑USENet was primarily evaluated on phantom data with quasi‑static compression, and its generalization to dynamic cardiac motion has not been demonstrated. In addition, RAFT‑USENet~\cite{majumder2025useraft} incorporates a tissue incompressibility constraint into the loss function, which assumes a fixed relationship between axial and lateral strains. This assumption may be violated in anisotropic tissues or complex cardiac motion in three dimensions. CrossRAFT adopts a bidirectional unsupervised training strategy that relies on SSIM-based photometric and smoothness losses, avoiding biomechanical priors and allowing the model to adapt to various tissue behaviors. A direct quantitative comparison with RAFT‑USENet is currently not possible because its code and model weights have not been publicly released.

Several limitations and future directions emerge from this work.
Regarding computational efficiency, Table~\ref{Tab_benchmark} shows that the inference time of CrossRAFT is 0.65s per pair on the simulated phantom dataset, higher than that of RAFT~\cite{teed2020raft} and MPWC++~\cite{tehrani2021mpwc++}, but it remains within clinically acceptable for real-time applications. This increase is attributed to two primary factors: the larger parameter count of the CrossRAFT architecture and the fact that current DL frameworks (e.g., PyTorch) are predominantly optimized for real-valued tensor operations. Consequently, the execution of complex-valued computations presents additional overhead, suggesting significant potential for speed optimization in future dedicated implementations. Future work will involve validation on \textit{in vivo} human datasets across various organs and pathological conditions to assess clinical robustness and generalization. Extending the CrossRAFT paradigm to volumetric ultrasound would be a natural progression, as out-of-plane motion remains a major source of decorrelation in 2D ultrasound imaging. 3D CrossRAFT can be evaluated on an open-access datasets, STRAUS~\cite{alessandrini20153Ddataset}, dedicated for 3D ultrasound strain imaging. Finally, exploring the integration of biomechanics-inspired constraints into the refinement loop could further regularize estimates in physiologically relevant ways, especially for structured tissues like myocardium.

\section{Conclusion}
\label{sec_conclu}
In this work, we proposed CrossRAFT, an end‑to‑end network for motion estimation directly from complex‑valued ultrasound signals (IQ and analytic data). By incorporating complex‑valued feature encoders and a custom correlation module, CrossRAFT extracts and exploits essential phase information that remains inaccessible to existing DL‑based motion estimation methods. The model adopts supervised pre‑training on the simulated phantom dataset, followed by bidirectional unsupervised learning with SSIM‑based and smoothness losses. Evaluated on simulated phantom, synthetic echocardiographic, and private \textit{in vivo} echocardiographic data, CrossRAFT delivers competitive accuracy on the phantom dataset and superior performance on the synthetic cardiac dataset compared to RF‑based baselines. These experimental results highlight the potential of the CrossRAFT framework to enhance downstream ultrasound applications, particularly in elastography and functional cardiac imaging. Future work will focus on extending the framework to three‑dimensional motion estimation and optimizing the architecture for real‑time deployment.

\section{Data and Code Availability}
The code of CrossRAFT will be made publicly available at: \url{https://github.com/Valeria-Leng/CrossRAFT.git}.

% \newpage
\bibliographystyle{IEEEtran}
\bibliography{Sample.bib}

\vfill
\end{document}